\documentclass[12pt,preprint]{aastex}
\usepackage{psfig,lscape}
\def\lap{\lower.5ex\hbox{$\; \buildrel < \over \sim \;$}}
\def\gap{\lower.5ex\hbox{$\; \buildrel > \over \sim \;$}}

\def\ergcm2s{${\rm erg\ cm^{-2}\ s^{-1}}$}

\def\ergscm2s{${\rm erg\ cm^{-2}\  s^{-1}}$}

\def\cm-2{${\rm cm^{-2}}$}
\def\ergs{${\rm erg\ s^{-1}}$}

\begin{document}

\title{Two New X-ray/Optical/Radio Supernova Remnants in M31}

\author{Benjamin F. Williams\altaffilmark{1},
Lor\'{a}nt~O.~Sjouwerman\altaffilmark{2},
Albert~K.~H.~Kong\altaffilmark{1}, Joseph~D.~Gelfand\altaffilmark{1},
Michael~R.~Garcia\altaffilmark{1}, Stephen~S.~Murray\altaffilmark{1}}

\altaffiltext{1}{Harvard-Smithsonian Center for Astrophysics, 60
Garden Street, Cambridge, MA 02138; \hbox{williams@head.cfa.harvard.edu};
\hbox{jgelfand@cfa.harvard.edu}; \hbox{akong@head.cfa.harvard.edu};
\hbox{garcia@head.cfa.harvard.edu}; \hbox{ssm@head.cfa.harvard.edu}}
\altaffiltext{2}{National Radio Astronomy Observatory, P.O. Box O, Socorro, NM 87801; \hbox{lsjouwer@aoc.nrao.edu}}


\begin{abstract}

We compare a deep (37 ks) {\it Chandra} ACIS-S image of the M31 bulge
to Local Group Survey narrow-band optical data and Very Large Array
(VLA) radio data of the same region.  Our precisely registered images
reveal two new optical shells with X-ray counterparts.  These shells
have sizes, [S~II]/H$\alpha$ flux ratios, and X-ray spectral
properties typical of supernova remnants (SNRs) with ages of
9$^{+3}_{-4}$ and 17$^{+6}_{-9}$ kyr.  Analysis of complementary VLA
data reveals the radio counterparts, further confirming that they are
SNRs.  We discuss and compare the properties and morphologies of these
SNRs at the different wavelengths.

\end{abstract}

\section{Introduction}

The high-spatial resolution of the {\it Chandra} X-ray Observatory is
allowing supernova remnants (SNRs) in M31 to be resolved at X-ray
wavelengths for the first time \citep{kong2002s,kong2003}.  These
detections provide the first opportunity to perform resolved
multi-wavelength studies of SNRs in M31.  Studies of these
extragalactic SNRs avoid several difficulties that hinder Galactic
studies, such as unreliable distances, large angular sizes, and high
Galactic absorption.  Multi-wavelength studies of these SNRs allow
reliable comparative size analyses for future determinations of
supernova rates and open the door to detailed studies of supernova
feedback in M31.

Optical emission-line surveys of M31 date back to \citet{rubin1972},
and X-ray surveys back to \citet{vanspeybroeck1979}.  Without digital
imaging to allow the subtraction of continuum emission from the
emission-line images and without high spatial resolution X-ray data,
reliable determination of counterparts was difficult for these early
surveys.  While optical surveys continued to catalog hundreds of SNRs
in M31 (e.g. \citealp{braun1993,magnier1995,williams1995}), ROSAT
studies had little luck in discovering X-ray SNRs in M31
\citep{magnier1997x}.

Very recently, with deep {\it Chandra} and {\it XMM} images of M31,
finding X-ray counterparts of optical SNRs has become more feasible.
\citet{kong2002s} found a previously-known optical SNR that was
well-resolved in {\it Chandra} images.  Later, \citet{kong2003} found
2 previously-unclassified SNRs that were resolved in X-ray, optical,
and radio images. 

Herein we report the discovery of 2 more SNRs in M31, found by
comparing narrow-band images from the Local Group Survey (LGS;
\citealp{massey2001}) to a precision-aligned, deep {\it Chandra}
image.  One object, CXOM31~J004248.9+412406 (r3-84), R.A.=00:42:48.97,
Dec.=+41:24:06.9 (J2000), coincides with one of the two SNR candidates
reported independently by \citet{trudolyubov2004} (XMMU
J004249.1+412407).  The second object, CXOM31~J004224.1+411733
(r2-57), R.A.=00:42:24.16, Dec.=+41:17:33.6 (J2000), also has the
X-ray, optical, and radio properties of an SNR and has not been
reported elsewhere.  Section~\ref{data} describes the data and
analysis techniques used.  Section~\ref{results} discusses the X-ray,
optical and radio properties of the SNRs, and
section~\ref{conclusions} provides a summary of our conclusions.

\section{Data Analysis}\label{data}

\subsection{Optical Data}

We obtained the [O~III], [S~II], H$\alpha$ and $V$-band images of the
LGS field 5 from the LGS
website.\footnote{http://www.lowell.edu/$\sim$massey/lgsurvey} These
images have already been properly flat-fielded and the geometric
distortions removed so that the coordinates in the images are good to
$\sim$0.25$''$ on the FK5 system and the images at the different
bandpasses are registered with one another.  We therefore were easily
able to subtract the $V$-band continuum from the [O~III] image and the
$R$-band continuum from the [S~II] and H$\alpha$ images in order to
make the line-emitting sources stand out.

We performed a rough calibration of the LGS [O~III] image by matching
the [O~III] fluxes of 10 planetary nebulae (PNe) with published
[O~III] fluxes \citep{ciardullo1989}.  This calibration provided a
conversion factor of 5.5 $\times 10^{-16}$ erg cm$^{-2}$ ct$^{-1}$.
We also roughly calibrated the H$\alpha$ and [S~II] images by matching
the fluxes of the SNR DDB 1-15 \citep{dodorico1980} to the fluxes
measured in the calibrated data set of \citet{williams1995} (H$\alpha
= 7.3 \times 10^{-14}$ \ergcm2s ; [S~II] = 5.5 $\times 10^{-14}$
\ergcm2s ).  This calibration yielded conversion factors of 1.0
$\times 10^{-16}$ erg cm$^{-2}$ ct$^{-1}$ and 1.8 $\times 10^{-16}$
erg cm$^{-2}$ ct$^{-1}$ in H$\alpha$ and [S II] respectively.  Using
these [O~III], H$\alpha$, and [S II] factors, we converted the LGS
count rates to units of \ergcm2s . 

\subsection{X-ray Data}

We also obtained a deep {\it Chandra} ACIS-S image centered on the M31
nucleus (ObsID 1575).  This data set, obtained on 05-Oct-2001, had an
exposure time of 37.7 ks, target R.A.=00:42:44.4, target
Dec.=41:16:08.3, and a roll angle of 180.42 degrees.  We created
exposure maps for this image using the CIAO script {\it
merge\_all},\footnote{http://cxc.harvard.edu/ciao/download/scripts/merge\_all.tar}
and we found and measured positions for the sources in the image using
the CIAO task {\it
wavdetect}.\footnote{http://cxc.harvard.edu/ciao3.0/download/doc/detect\_html\_manual/Manual.html}
This processing identified sources in the image down to a flux limit
of $\sim$8 $\times$10$^{-16}$ \ergcm2s assuming an absorbed power-law
spectrum with slope 1.7 and $N_H = 10^{21}$ cm$^2$, or a (unabsorbed)
luminosity limit of $\sim$7 $\times$10$^{34}$ \ergs\ in M31, assuming
a distance of 780 kpc \citep{stanek1998,williams2003b}.

\subsubsection{X-ray/optical Image Alignment}

We aligned the coordinate system of the ACIS-S image with the LGS
coordinate system by translating and adjusting the plate scale of the
ACIS-S coordinate system so that 13 globular cluster sources had the
same coordinates as the centroids of the respective globular clusters
in the LGS $V$-band image.  This transformation, performed using the
IRAF\footnote{IRAF is distributed by the National Optical Astronomy
Observatory, which is operated by the Association of Universities for
Research in Astronomy, Inc., under cooperative agreement with the
National Science Foundation.} task {\it ccmap}, had root-mean-square
(rms) residuals of 0.16$''$ in RA and 0.15$''$ in DEC.  These errors
were added in quadrature to the position errors of the sources
determined by {\it wavdetect} to determine the final position errors
for X-ray sources on the LGS [O III] image.  

\subsubsection{X-ray spectra}

We extracted energy spectra and the associated responses from the
ACIS-S image with the CIAO task {\it
psextract}\footnote{http://cxc.harvard.edu/ciao/ahelp/psextract.html}
and CALDB version 2.26, which automatically corrects for the
degradation in the effective low-energy quantum efficiency of the ACIS
detectors. The background spectra were extracted with an annulus
region.  Since we only have 43 counts (37 background-subtracted) for
r2-57 and 37 counts (31 background-subtracted) for r3-84, we used two
methods to fit the spectra from 0.3--5 keV.

We first binned the spectra with $>$5 counts per bin and employed
$\chi^2$-Gehrels statistics to find the best fit using the CIAO
3.0/Sherpa fitting package \citep{freeman2001}.  Errors were estimated
using the Sherpa command {\it
projection},\footnote{http://cxc.harvard.edu/ciao/ahelp/projection.html}
which varies each parameter's value independently along a grid of
values to determine the 1$\sigma$ confidence intervals.

We fit the spectra with two single-component models with absorption:
power-law and Raymond-Smith (RS).  The power-law model is commonly a
good fit to the continuum of X-ray spectra and has 3 free parameters:
index, absorption column, and normalization.  The RS model is often
used to describe the spectra of SNRs.  We applied this model with the
abundance parameter fixed to solar and the redshift fixed to zero,
leaving 3 free parameters: temperature, absorption column and
normalization.  Results are discussed in \S~\ref{xspec} and shown in
Figure~\ref{spec}.  To verify the results, we then fit the unbinned
spectra (background not subtracted) with Cash statistics
\citep{cash1979}.

We further investigated the nature of the X-ray sources by calculating
the hardness ratios of the detected counts.  These were calculated
from counts extracted in 3 different energy bins.  The soft bin (S)
contains photons of energies 0.3--1 keV.  The medium bin (M) contains
photons of energies 1--2 keV.  The hard bin (H) contains photons of
energies 2--8 keV.  The final background-subtracted hardness ratio
equations were H1 = (M-S)/(H+M+S) and H2 = (H-M)/(H+M+S)
\citep{prestwich2003}.  These hardness ratios are discussed in
\S~\ref{xspec}.

Finally, we checked the results of the {\it Chandra} energy spectra
with {\it XMM-Newton} archival data, since the effective area of {\it
XMM-Newton} is much larger than that of {\it Chandra}.  There are four
{\it XMM-Newton} observations of the center of M31 with exposure times
from $\sim$10--60 ks.  Unfortunately, one of the SNRs (r2-57; see
\S~\ref{search}) lies too close to the center of M31 to be resolved
from the diffuse emission at the resolution of {\it XMM-Newton}; we
therefore could not perform a reliable spectral fit for any {\it
XMM-Newton} observation of this SNR.

The other SNR (r3-84; see \S~\ref{search}) is in the gap between CCD
chips in two of the four {\it XMM-Newton} observations.  We performed
spectral analyses on the other two observations, taken on 29-Jun-2001
(ObsID=0109270101, target R.A.=00:42:43, target Dec.=41:15:46,
P.A.=76.05 degrees, medium filter, and exposure=30.6 ks) and
06-Jan-2002 (ObsID=0112570101, target R.A.=00:42:43, target
Dec.=41:15:46, P.A.=249.84 degrees, thin1 filter, and exposure=56.5
ks). We extracted the 0.3--5 keV spectra with the {\it XMM-Newton} SAS
package v5.4.1; only EPIC-pn spectra were considered because of the
higher sensitivity.

The spectra were binned to have at least 15 counts per spectral bin in
order to allow the use of $\chi^2$ statistics.  Background spectra
were extracted from source free regions.  The source contained 150
counts (84 background-subtracted) in the 27 ks observation and 314
counts (205 background-subtracted) in the 51 ks observation.  Results
are discussed in \S~\ref{xspec}, and shown in Figure~\ref{xmm}.
 
\subsubsection{X-ray PSF Tests}\label{psftests}

As both of the newly-discovered SNRs are several arcminutes off-axis
in the {\it Chandra} data, there is a possibility that the larger
off-axis point spread function (PSF) of {\it Chandra} could be
misinterpreted as an extended source in the X-ray image.  To test this
possibility, we used the SNRs' spectral fits to simulate their {\it
Chandra} PSF at their location on the ACIS-S detector.  The simulation
was performed with the web-based PSF simulator
ChaRT\footnote{http://cxc.harvard.edu/chart/}.

\begin{figure}
\centerline{\psfig{file=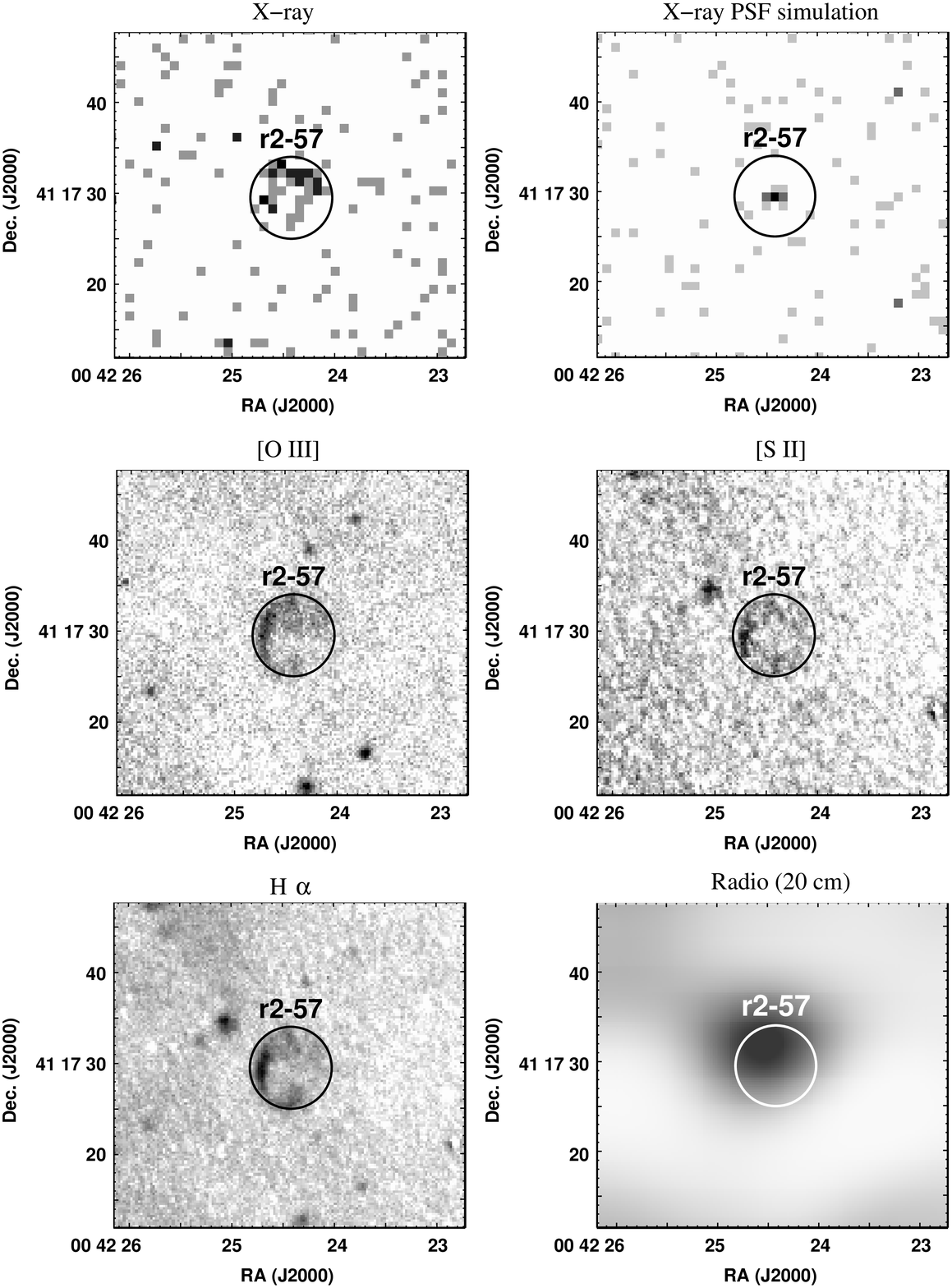,height=7in,angle=0}}
\caption{The new SNR r2-57 is shown at 5 wavelengths, X-ray,
H$\alpha$, [O~III], [S~II], and low-resolution radio (20 cm).  The
circles (9$''$ diameter, $\sim$34 pc) show the approximate size of the
SNR.  The X-ray PSF has FWHM = 1.4$''$ in this image, showing the
r2-57 is clearly resolved in X-rays.  For comparison, a simulation of
the {\it Chandra} PSF at this location in the focal plane is shown
(upper right).  The radio image is from the low resolution 20 cm
archive VLA data. The image has too low resolution ($\sim$ 13$''$) to
allow a detailed comparison with the optical and X-ray morphologies.
The deconvolved size is given in the text.}
\label{r257}
\end{figure}

\begin{figure}
\centerline{\psfig{file=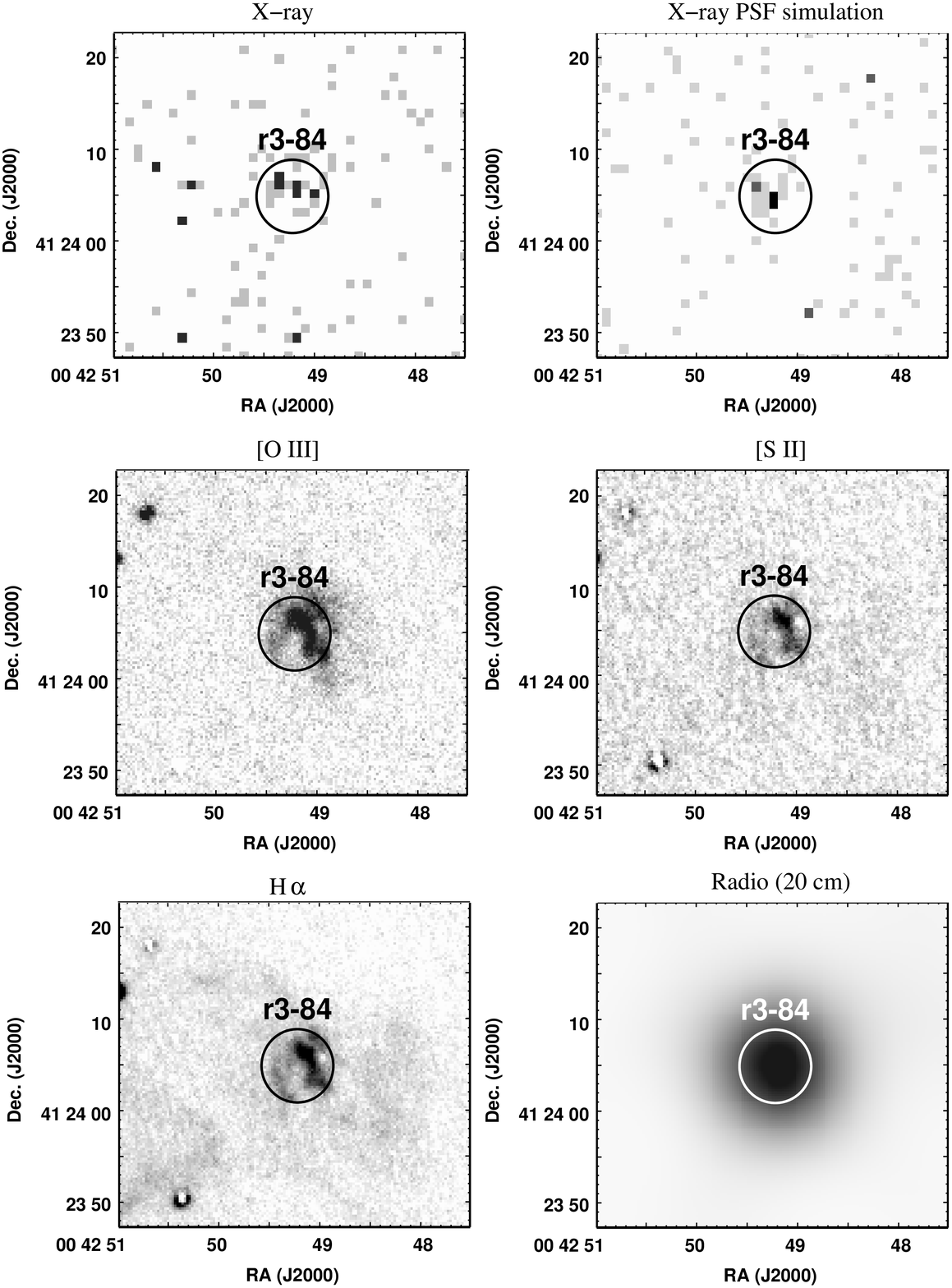,height=7in,angle=0}}
\caption{The new SNR r3-84 is shown at 5 wavelengths, X-ray,
H$\alpha$, [O~III], [S~II], and low-resolution radio (20 cm).  The
circles (8$''$ diameter, $\sim$30 pc) show the approximate size of the
SNR.  The X-ray PSF has FWHM = 4.1$''$ in this image, showing the
r3-84 is only marginally resolved in X-rays.  For comparison, a
simulation of the {\it Chandra} PSF at this location in the focal
plane is shown (upper right).  The radio image is from the low
resolution 20 cm archive VLA data. The image has too low resolution
($\sim$ 13$''$) to allow a detailed comparison with the optical and
X-ray morphologies.  The deconvolved size is given in the text.}
\label{r384}
\end{figure}

\begin{figure}
\centerline{\psfig{file=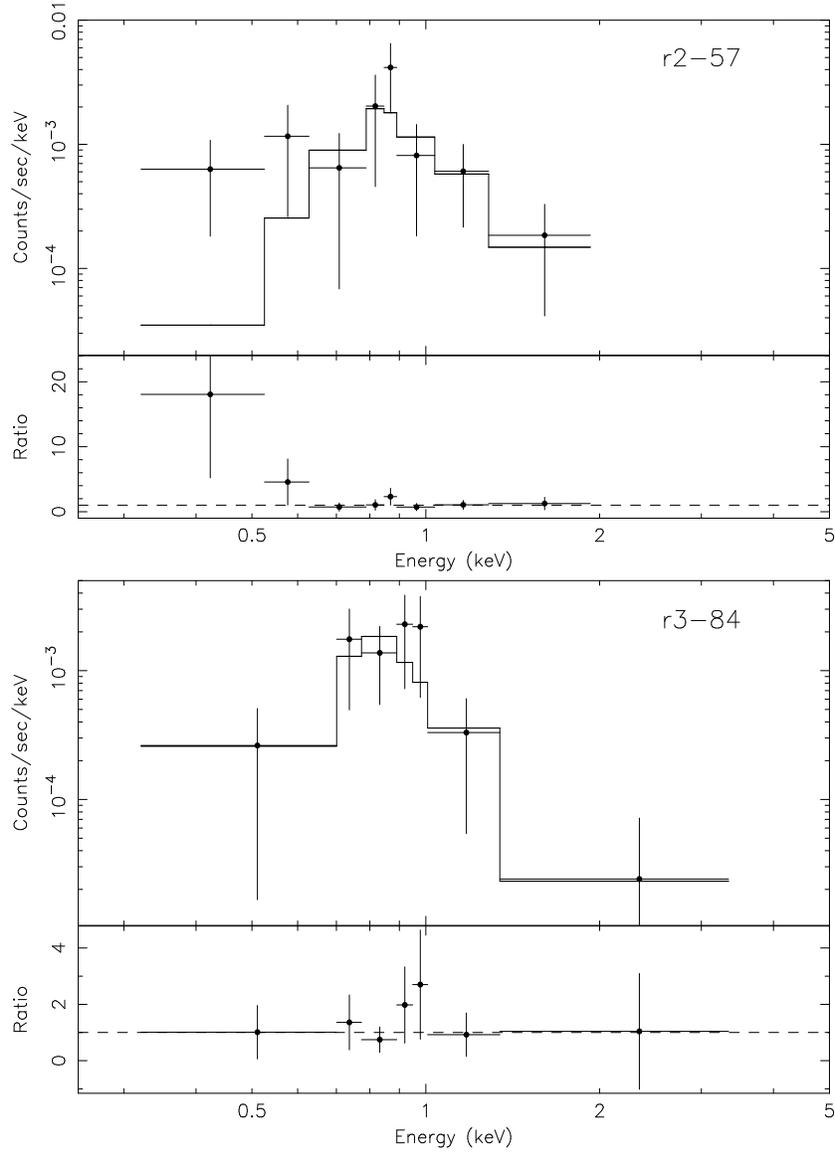,height=6.0in,angle=0}}
\caption{{\it Chandra} spectra of r2-57 and r3-84 from OBSID 1575. The
best-fit absorbed Raymond-Smith models are shown with histograms on
each plot.  Source r2-57 fit (top) parameters are
$N_H=8.9\times10^{21}$ cm$^{-2}$ and kT$_{RS}=0.17$ keV.  Source r3-84
fit (bottom) parameters are $N_H=4\times10^{21}$ cm$^{-2}$ and
kT$_{RS}=0.3$ keV.}
\label{spec}
\end{figure}

PSF simulations containing about the same number of counts as the
detections are shown next to the X-ray SNR detections in
Figures~\ref{r257} and \ref{r384}.  We also produced PSF simulations
with $\sim$10$^5$ counts.  Azimuthally averaged profiles of these
simulations and of the SNR detections were measured in 1.5$''$ annuli.
$\chi^2$ tests of the detected SNRs' profiles against the simulated
PSFs (normalized to the surface brightness of the central 3$''$ of the
SNR detection) were calculated to assess whether the SNRs were
resolved (results in \S~\ref{xraysize}).  The number of counts in each
annulus ranged from 1--21 in the SNR detections.  Profile errors were
determined by Gehrels statistics \citep{gehrels1986} in annuli with
$\leq$10 counts and by standard Poisson statistics in annuli
containing $>$10 counts.  Results are discussed in \S~\ref{xraysize}.

\subsection{Radio Data}\label{radiodata}

The radio data was collected from a 20 cm VLA survey source list
presented by Braun (1990) complemented by a recent 6 cm VLA B-array
observation \citep{sjouwerman2004}. We also retrieved 20 cm low
resolution VLA C-array and 6 cm VLA D-array data from the NRAO
archive. These data sets, albeit pointed to the center of M31 instead
of directly at the SNRs, were the most sensitive data sets to search
for arcsecond extended radio counterparts.

\citet{sjouwerman2004} observed a field-of-view of about 8$'$ using
the VLA at 6 cm for 22 hours on three days in June 2002, obtaining an
angular resolution of 1.2$''$ and the ability to detect angular
structures up to about 35$''$. The 20 cm (6 cm) archive VLA data with
a field-of-view of about 30$'$ (8$'$) in diameter was taken during 20
(19) hours on three (four) days in August 1993 (July 1992), has an
angular resolution of 13$''$ (14$''$) and is sensitive to angular
structures up to about 15$'$ (5$'$).  The data were all calibrated and
imaged in NRAO's AIPS package using the new VLARUN pipeline procedure
with additional self-calibration for the 20 cm archive data as
outlined in the AIPS Cookbook. The resulting rms noise in the images
is 60 (15) $\mu$Jy for the 20 cm (6 cm) archive data, and 6 $\mu$Jy
for the 6 cm \citet{sjouwerman2004} data.

Finally, we retrieved 20 cm high resolution (1$''$; A-array) VLA data
from the NRAO archive to detect any compact components of these SNR
candidates.  These data consist of eleven pointings in the center and
southern half of M31.  The data were also calibrated and imaged with
the VLARUN pipeline procedure.  Images with an rms noise of
$\sim$0.2~mJy were produced.

\subsection{SNR Search}\label{search}

We visually searched the aligned [O III] image for shell structures
with X-ray counterparts by placing 2$''$ radius circles onto the
[O~III] image centered on the locations of all X-ray sources detected
in the ACIS-S images.  This search yielded 5 counterpart candidates, 3
of which have been cataloged as X-ray SNR and were previously
discussed in \citet{kong2002s,kong2003}.  The X-ray images were then
inspected at the locations of the two new SNRs.  Figures~\ref{r257}
and \ref{r384} show the {\it Chandra}, H$\alpha$, [O~III], [S~II], and
VLA (low resolution 20 cm) images of each new SNR.  These 2 new
matches showed previously-cataloged X-ray counterparts:
CXOM31~J004224.1+411733 (r2-57) and CXOM31~J004248.9+412406 (r3-84)
\citep{kong2002}.

Once the two new matches in optical and X-ray were found, radio
counterparts were sought for in the list of Braun (1990). Indeed,
r3-84 is listed as an extended source (source number 97).  Source
r2-57 is located near an area of diffuse emission and, although not
listed in Braun (1990), is marginally visible in his Figure 3. We
therefore also checked our recent 6 cm VLA data
\citep{sjouwerman2004}. The 6 cm data has only a limited
field-of-view, with a full-width at half-maximum (FWHM) of 500$''$,
meaning that r3-84 is too far away from the field center (8$'$) to be
visible. However we found a $\sim$10$''$ extended 2$\sigma$ patch at
the position of r2-57. This triggered a search in the VLA archive for
more sensitive data for which we chose the sets discussed in
\S~\ref{radiodata}. The archive data does confirm a radio detection of
r2-57, at 20 cm and at 6 cm, and yields an additional independent
radio detection of r3-84 at 20 cm. No other radio matches were found
in the literature.

\section{Results}\label{results}

\subsection{X-ray Properties}

\subsubsection{Spectral Properties}\label{xspec}

\begin{figure}
\centerline{\psfig{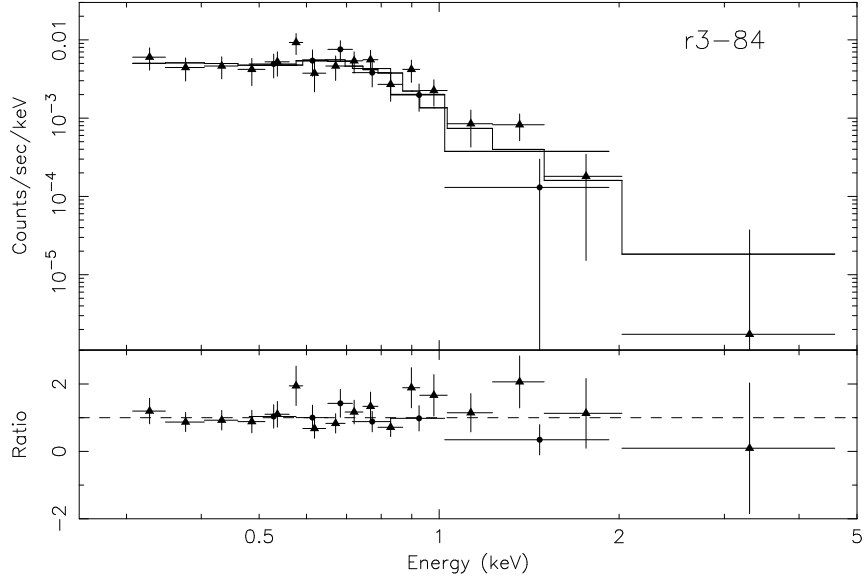}}
\caption{{\it XMM-Newton} EPIC-pn spectrum of r3-84. The fit shown by
the histogram is a combination Raymond-Smith and power-law model
($N_H=7\times10^{20}$ cm$^{-2}$, kT$_{RS}=0.25$ keV, and
$\alpha=3.33$) and the unabsorbed 0.3--7 keV luminosity is
$1.7\times10^{36}$ erg s$^{-1}$. Data taken on 2001 June 29 are marked
as circles, while triangles are data taken on 2002 January 6.}
\label{xmm}
\end{figure}

We were able to constrain the physical properties of both SNRs with
their spectra from {\it Chandra} and {\it XMM-Newton}.  Plots of the
best fits using the RS model are provided in Figure~\ref{spec} and
\ref{xmm}.

For r2-57, the binned spectrum is well-fit by a power-law model with
$\alpha=5.3^{+0.6}_{-0.8}$~(1$\sigma$),
$N_H=2.8^{+0.8}_{-0.5}\times10^{21}$ cm$^{-2}$ ($\chi^2_{\nu}=0.6$ for
5 degrees of freedom (dof)) and 0.3--7 keV luminosity of
3.4$^{+0.5}_{-0.4}\times10^{37}$ erg~s$^{-1}$. The spectrum is also
fit well by an RS model with kT~=~0.17$^{+0.54}_{-0.06}$ keV and
$N_H$~=~(8.9$\pm$3.2)~$\times10^{21}$ cm$^{-2}$ ($\chi^2_{\nu}=0.9$
for 5 dof) and an absorption-corrected 0.3--7 keV luminosity of
4.3$^{+1.9}_{-3.8}\times10^{37}$~erg~s$^{-1}$.  Fits of the unbinned
spectrum give similar results.

In the case of r3-84, a power-law model does not fit the binned
spectrum with reasonable parameters; the photon index is $>$10
suggesting a soft X-ray source. On the other hand, the spectrum is
well fit by an RS model with kT $=0.3^{+0.5}_{-0.1}$ keV,
$N_H$~=~4$^{+17}_{-4}\times10^{21}$ cm$^{-2}$ ($\chi^2_{\nu}=0.4$ for
4 dof) and an absorption-corrected 0.3--7 keV luminosity of
2.3$^{+1.1}_{-1.2}\times10^{36}$ erg s$^{-1}$.

We checked our {\it Chandra} spectral results using the two {\it
XMM-Newton} spectra of r3-84.  These spectra were fit with several
single-component models (with absorption) including power-law, RS,
non-equilibrium ionization (NEI), and bremsstrahlung models.  
The NEI and bremsstrahlung models give acceptable fits.  The (solar
abundance) NEI model has $\chi^2_{\nu}=1.11$ for 19 dof with
$N_H=3.4^{+0.9}_{-2.0}\times10^{21}$ cm$^{-2}$,
kT$=0.45^{+0.51}_{-0.08}$ keV, and ionization age $\tau =
(1.2\pm0.2)\times10^{10}$ s cm$^{-3}$; the modeled unabsorbed 0.3--7
keV luminosity is $(9\pm1)\times10^{36}$ erg s$^{-1}$.  The
best-fitting single-component model is bremsstrahlung; it has
$\chi^2_{\nu}=0.97$ for 20 dof with $N_H=(2.9\pm2.4)\times10^{21}$
cm$^{-2}$ and kT$_{brem}=0.2\pm0.1$ keV; the modeled unabsorbed 0.3--7
keV luminosity is (9.4$\pm$2.5)~$\times10^{36}$ erg s$^{-1}$.

The overall best-fitting model for the {\it XMM} spectra is an RS plus
power-law\footnote{The single component RS model had $N_H <
1.3\times10^{21}$, kT$_{RS}=0.25\pm0.03$ keV, and modeled unabsorbed
0.3--7 keV luminosity (7$\pm$3)$\times10^{35}$ erg s$^{-1}$, but it
had a poor $\chi^2_{\nu}$ (1.44 for 20 dof).} (see Figure~\ref{xmm}).
It provides a good fit ($\chi^2_{\nu}=0.86$ for 18 dof) with
$N_H=7^{+2}_{-7}\times10^{20}$ cm$^{-2}$,
kT$_{RS}=0.25^{+0.04}_{-0.06}$, and power-law slope
$\alpha=3.3^{+1.4}_{-1.0}$; the modeled unabsorbed 0.3--7 keV
luminosity is $1.7^{+0.3}_{-0.7}\times10^{36}$ erg s$^{-1}$.  These
results are consistent with the {\it Chandra} results as well as the
temperature and luminosity ranges determined from fits to the {\it
XMM} data by \citet{trudolyubov2004}, but the temperature and
luminosity are both significantly lower than those given by the NEI
model fit.  We use the {\it Chandra} temperature results, which are
consistent with all of the {\it XMM} fits, to calculate age and
density estimates in \S~\ref{multiprops}.

While these spectral fits do not allow detailed modeling of the
sources, they show that the X-ray spectra and luminosities are typical
of X-ray SNRs.  The softness of the spectra is confirmed by the
hardness ratios.  For r2-57, H1~=~$-$0.5$\pm$0.2, and
H2~=~$-$0.4$\pm$0.1.  For r3-84, \hbox{H1~=~$-$0.6$\pm$0.2}, and
H2~=~$-$0.2$\pm$0.1; these ratios are typical of SNRs
\citep{prestwich2003}.  Neither of these sources contain counts with
energies higher than 2 keV, also consistent with typical X-ray SNRs.

\subsubsection{Variability}

The variability of these X-ray sources has been investigated by
\citet{kong2002}, who found r2-57 to have constant flux.  On the other
hand, r3-84 was classified as a variable source in their survey,
inconsistent with our finding that this source is an SNR.  We checked
the variability of this source, and it is variable according to their
criteria due to a low count rate in one detection.  Further inspection
of the detection with the low count rate shows that r3-84 is on the
edge of the chip, so that some of the source flux could have missed
the detector, calling into question the classification of this source
as variable.

\subsubsection{X-ray Sizes}\label{xraysize}

In addition to the spectrophotometric properties of the SNRs, the {\it
Chandra} data allow estimates of the SNRs' sizes.  The X-ray detection
of r3-84 appears as a very faint shell structure 6$''$ across.  Though
the structure appears to be this size in the detection, the {\it
Chandra} PSF 8$'$ off axis (where this detection of r3-84 is located)
has a FWHM of 4.1$''$ according to our ChaRT simulation
(see~\S~\ref{psftests}).  A $\chi^2$ comparison of the azimuthally
averaged profile of our detection of r3-84 against the azimuthally
averaged simulated PSF at the same location in the {\it Chandra} focal
plane yields a $\chi^2/\nu$ of 9.2/4, leaving only a 5.6\% chance that
this object is a point source.  The X-ray simulation, X-ray image,
optical images, and radio image of the source are shown on the same
scale, with circles of 8$''$ diameter in Figure~\ref{r384}.  The X-ray
size is very similar to that of the optical (and the low resolution
radio) counterpart.

The X-ray detection of r2-57 also appears as a very faint shell
structure, but at $\sim$8.5$''$ across and 4$'$ off-axis (where the
ACIS-S point spread function is 1.4$''$ according to our ChaRT
simulation), this source is clearly resolved.  The PSF simulation
(see~\S~\ref{psftests}) shows no similarity to the shell-like source
detection; a $\chi^2$ comparison of the azimuthally averaged profile
of our detection of r2-57 against the azimuthally averaged simulated
PSF at the same location in the {\it Chandra} focal plane yields a
$\chi^2/\nu$ of 28.1/4, leaving a probability of 10$^{-5}$ that this
object is a point source.  The X-ray simulation, X-ray image, optical
images, and radio image of the source are shown on the same scale,
with circles of 9$''$ diameter in Figure~\ref{r257}.

\subsection{Optical Properties}

The narrow-band LGS images provided estimates of the H$\alpha$,
[S~II], and [O~III] luminosities of these SNR, as well as the SNR
sizes.  The relative strength of these emission lines is a well-known
diagnostic for distinguishing shock-heated SNRs from photo-ionized
H~II regions and planetary nebulae (e.g. \citealp{levenson1995}).  In
addition, the SNR sizes and structures help to constrain their ages
(e.g. \citealp{kong2002s}).

We measured the counts in the narrow-band images using an aperture of
4$''$ radius centered on r3-84 and an aperture of 5$''$ radius
centered on r2-57.  The measured H$\alpha$, [S~II], and [O~III]
fluxes, in units of 10$^{-14}$ erg cm$^2$ s$^{-1}$ were 1.0, 0.9, and
1.5 for r2-57, respectively, and 1.1, 0.9, and 1.7 for r3-84,
respectively.  These [S~II]/H$\alpha$ ratios of $\sim$0.9 are strong
indications that these objects are SNRs.  Typically, SNRs have
[S~II]/H$\alpha > 0.4$, while the ratios of H II regions are lower
\citep{levenson1995}.

The [O~III]/H$\alpha$ ratio must be corrected for absorption because
of the large difference in wavelength.  We used the $N_H$ values from
the X-ray spectral fits (see \S~\ref{xspec}) to estimate the reddening
values of the SNRs \citep{predehl1995}.  These values were $A_V \sim
A_{[O~III]} \sim 5.0$ for r2-57 and $A_V \sim A_{[O~III]} \sim 2.2$
for r3-84.  Assuming a standard reddening law, $A_R \sim A_{H\alpha}
\sim 3.7$ for r2-57 and $A_R \sim A_{H\alpha} \sim 1.6$ for r3-84, so
that the absorption-corrected [O~III]/H$\alpha$ ratios are factors of
3.2 and 1.7 higher than the measured ratios, for r2-57 and r3-84,
respectively.  The [O~III]/H$\alpha$ ratios are therefore $\sim$5 and
$\sim$3 for r2-57 and r3-84, respectively.  Such ratios are slightly
high, but not unprecedented for SNRs containing a variety of shock
velocities $\gap$100 km s$^{-1}$
(e.g. \citealp{vancura1992,fesen1997,mavromatakis2000}).

Finally, we measured the sizes of r2-57 and r3-84 using the ruler
option in the image viewing program {\it ds9}.  The sizes were
measured in all three bandpasses.  The H$\alpha$ sizes were 8.8$''$
and 6.6$''$ respectively. The [S II] sizes were 8.4$''$ and 6.4$''$
respectively, and the [O III] sizes were 8.8$''$ and 6.2$''$,
respectively.  Combining these measurements yields optical sizes of
8.7$''\pm$0.2 (33$\pm$1 pc) and 6.4$''\pm$0.2 (24$\pm$1 pc),
respectively.  These sizes were used to calculate our age and density
estimates in \S~\ref{multiprops}.

\subsection{Radio Properties}

Unfortunately the most sensitive radio data available has insufficient
resolution and/or sensitivity to obtain more detailed morphologies
than the deconvolved sizes.  

The deconvolved size of r2-57 at 20 cm is 4.3$''\pm$2.8, and
$\sim$10$''\pm$9 at 6 cm.  Because r2-57 has been detected at two
different wavelengths (20 and 6 cm) with similar ($u,v$)-coverage, a
radio spectral index could be obtained from the integrated flux
densities, about 0.25$\pm$0.09 and 0.24$\pm$0.22 mJy at 20 and 6 cm,
respectively. Unfortunately, a spectral index taken from such
measurements would be unreliable considering the large errors.  These
large errors are attributable to the weakness of the signals as well
as the (unreliable) correction factor for r2-57 at 6 cm due to loss of
sensitivity near the half-power point of the antennas. Nevertheless,
we conclude that the radio spectral index for r2-57 is not
inconsistent with average spectral indices of SNRs.

In the 20 cm VLA archive data we measured an integrated flux density
of 1.2$\pm$0.2 mJy and deconvolved size of 8.6$''\pm$1.5 for r3-84. In
the 20 cm high-resolution archival data, a point source of 0.8 mJy is
seen coincident with the brightest region in the lower-resolution
data.  Because r3-84 is too far away from the phase center in the 6 cm
data sets to be detected, we are unable to comment on a spectral index
for this source.

\subsection{Comparing X-ray/Optical/Radio Morphologies}\label{multiprops}

The similar size measurements for these SNRs at X-ray, optical, and
radio wavelengths suggests at first that these SNRs have similar
morphologies at all wavelengths.  However, in the case of r2-57, the
brightest region in all three optical wavelengths is the eastern rim,
while the brightest regions in the radio and X-ray images are the
north-northeast and northwest parts of the SNR, respectively.  Even
with this low-count X-ray detection, we can begin to learn about the
multi-wavelength structure of this SNR.

\begin{figure}
\centerline{\psfig{file=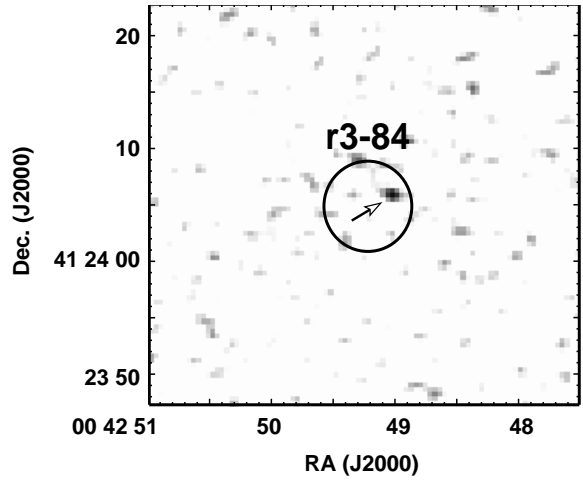,height=2.5in,angle=0}}
\caption{A high-resolution radio image (20 cm) of r3-84 is shown at
1$''$ resolution.  The circle (8$''$ diameter, $\sim$30 pc) shows the
approximate size and position of the SNR.  The arrow indicates a
bright radio region detected on the northwest edge of the SNR, the
location of the brightest optical and X-ray emission.  Deeper, lower
resolution radio data, shown in Figure~\ref{r384}, suggest that a
larger radio shell is present, but too faint to be detected in this
high-resolution image.}
\label{r384r}
\end{figure}

Although the low spatial resolution of the deep radio data limits
detailed morphological comparison with the shorter wavelengths, the
deconvolved sizes of the SNRs provide some interesting comparisons.
The sizes are all consistent within the errors, but the size of r2-57
at 20 cm may be about half the size at 6 cm.  If there is a real size
difference, we attribute it as likely due to strong 20 cm radiation
originating in different areas of the SNR than the 6 cm radiation.
The small scale 20 cm structure seen in r3-84 (see Figure~\ref{r384r})
hints that this hypothesis is reasonable by showing how these SNRs can
have some complex structures at radio wavelengths.

The morphology of r3-84 appears to be more consistent across all
wavelengths.  While r3-84 is too far off axis and has too few counts
to say anything detailed about its X-ray morphology, there is a hint
that the northwest portion of the SNR is the brightest region in
X-rays.  This hint is consistent with the optical SNR, which is
brightest in the northwest portion in all 3 narrow-band images.  In
addition, our high-resolution 20 cm data, shown in Figure~\ref{r384r},
reveal that the northwest rim of the SNR houses the brightest radio
region, consistent with the X-ray/optical shell.

The deconvolved size of r3-84 is consistent with the
$\sim$10.0$''\times4.4''$ size given by Braun (1990) and the 8$''$
circles shown in Figure~\ref{r384}.  This extended emission is not
detected in the 1$''$ resolution data because the noise level in the
high-resolution data ($\sim$0.1 mJy/beam) is too high.  If the 1.2 mJy
source detected in the 20 cm low resolution VLA archive data is
homogeneously spread over the 8$''$ disk, then the average surface
flux density would be $\sim$0.04 mJy/beam in a 1.4$''$ beam.  Such
emission would be below the noise of the high-resolution data.  The
radio counterpart of this SNR is comprised of a $\sim$10$''$ shell
with a bright region in the northwestern portion.  Therefore, the
morphology of this SNR, unlike that of r2-57, appears quite similar at
all wavelengths.

As the measured sizes of the SNRs are consistent across all of the
observed wavelengths, we can use the sizes and measured temperatures
to approximate the ages using the Sedov solution as done in
\citet{kong2002s}.  We assume the SNRs are in the adiabatic expansion
phase.

We adopt radii of 17$\pm$1 pc (r2-57) and 12$\pm$1 pc (r3-84) and
shock temperatures of $T_s=0.17^{+0.54}_{-0.06}$~keV (r2-57) and
$T_s=0.3^{+0.5}_{-0.1}$ keV (r3-84).  We assume an initial explosion
energy of $E_{51}=0.3^{+0.4}_{-0.1}$ (in units of 10$^{51}$ erg
s$^{-1}$) for both SNRs.  This value of $E_{51}$ is the average value
from a sample of M31 SNRs surveyed by spectroscopic observations,
bracketed by the range of measured $E_{51}$ values, ignoring the
highest and lowest values \citep{blair1981}. From equations (1) and
(2) of \citet{kong2002s}, we obtain an age estimate of
$17^{+6}_{-9}$~kyr and a density estimate of
$n_0=0.2^{+0.7}_{-0.2}$~cm$^{-3}$ for r2-57.  Applying the same
calculations to r3-84 yields an age estimate of 9$^{+3}_{-4}$~kyr and
a density estimate of $n_0=0.3^{+1.2}_{-0.2}$~cm$^{-3}$.

\section{Conclusions}\label{conclusions}

By comparing {\it Chandra} ACIS-S images with Local Group Survey
[O~III] images, we have discovered 2 new X-ray/optical/radio SNRs in
M31.  Both of these objects are previously-cataloged X-ray sources.
One is a previously-cataloged radio source, and the other is a
newly-discovered radio counterpart discovered in our study.  One SNR
(r2-57) is well resolved in the X-ray images; the other is marginally
resolved.  These SNRs have emission-line ratios, X-ray spectra, and
sizes typical of other SNRs in M31. The temperatures and sizes of the
SNRs provide age estimates of $17^{+6}_{-9}$~kyr (r2-57) and
9$^{+3}_{-4}$~kyr (r3-84).

While these relatively weak X-ray detections have been crucial for the
discovery of resolved X-ray SNRs in M31, they have only begun to allow
studies of the multi-wavelength structure of these objects.  With a
total of 5 SNRs resolved by {\it Chandra} in X-rays so far, it seems
clear that, as deeper X-ray data become available, more of these
objects are likely to be discovered and more detailed multi-wavelength
morphological and spectral studies can be performed.

We thank John Raymond for advising in the interpretation of the
[O~III] fluxes.  Support for this work was provided by NASA through
grant number GO-9087 from the Space Telescope Science Institute and
through grant number GO-3103X from the {\it Chandra} X-ray Center.
MRG acknowledges support from NASA LTSA grant NAG5-10889.  The
National Radio Astronomy Observatory is a facility of the National
Science Foundation operated under cooperative agreement by Associated
Universities, Inc.

\end{document}